Title: Slow relaxation of magnetoresistance in doped p -GaAs/AlGaAs layers with partially filled upper Hubbard band.

Article Type: Communication

Section/Category:

Corresponding Author: N.V. Nina Agrinskaya,

Corresponding Author's Institution: Ioffe Phys.Techn.Institute

First Author: Nina Agrinskaya

Order of Authors: Nina Agrinskaya; V. I Kozub; D. V Shamshur; A. Shumilin

Manuscript Region of Origin:

Abstract: We observed slow relaxation of magnetoresistance in quantum well structures GaAs-AlGaAs with a selective doping of both wells and barrier regions which allowed partial filling of the upper Hubbard band. Such a behavior is explained as related to magnetic-field driven redistribution of the carriers between sites with different occupation numbers due to spin correlation on the doubly occupied centers. Such redistribution, in its turn, leads to slow multi-particle relaxations in the Coulomb glass formed by the charged centers

# Slow relaxation of magnetoresistance in doped p -GaAs/AlGaAs layers with partially filled upper Hubbard band.

N.V.Agrinskaya, V.I.Kozub, D.V.Shamshur and A.Shumilin

*A.F.Ioffe Physico-Technical Institute, St.-Petersburg, Russia*

**Abstract**

We observed slow relaxation of magnetoresistance in quantum well structures GaAs-AlGaAs with a selective doping of both wells and barrier regions which allowed partial filling of the upper Hubbard band. Such a behavior is explained as related to magnetic-field driven redistribution of the carriers between sites with different occupation numbers due to spin correlation on the doubly occupied centers. This redistribution, in its turn, leads to slow multi-particle relaxations in the Coulomb glass formed by the charged centers.

## 1. Introduction

During last years a concept of the Coulomb glass - the system of localized charged centers where interaction effects are as strong as the disorder ones - was widely discussed. In particular, it was shown numerically that such a glass, as the spin glasses, has a huge (supposedly exponentially large) number of pseudoground states ("valleys") which are close in energies and have very small intervalley transition rates [1]. Numerous experimental studies (see e.g. see, e.g. [2-5]) evidenced that the Coulomb glasses demonstrate interesting and non-trivial slow relaxation and memory effects. In our recent paper [6] we suggested an explanation of such a behavior basing on the idea of electronic polaron similar to the one suggested in Refs. [7]. We assumed that in the Coulomb glass there exist local multistable aggregates of sites with moderate transition rates corresponding to the experimentally observed time scale which form the polarons. Earlier we exploited the idea of the multistable aggregates to explain low-frequency noise in doped conductors [8]. However, to the best of our knowledge the memory and slow relaxation effects were mostly observed in complex systems like indium oxides (Refs. [2-5]) with strong degree of disorder. At the same time the standard doped semiconductors have not exhibited similar dynamics. One can expect that a particular reason is the usage of the gate voltage for manipulations of the systems under studies. Indeed, the variation of the gate voltage produces in particular large systematic effect on the density of states which can prevent an observation of the effects of slow dynamics if the latter are weak.

In the present paper we report an observation of slow relaxation of magnetoresistance in standard multiple well structures where both wells and barriers were selectively doped by acceptors. In these structures we realized a partial occupation of the upper Hubbard band of the centers $A^+$ within the wells which coexisted with the lower Hubbard band formed by the holes which can still be coupled to the acceptors within the barrier ($A^0$ and $A^-$ ). We believe that the external magnetic field leads to rearrangement of the charge distribution between the different centers and affects the slow relaxing aggregates and, correspondingly, the polaron gaps. The theoretical considerations given below are at least in a qualitative agreement with experimental results.

## 2. Experimental

The structures contain ten quantum GaAs wells of thickness 15 nm, separated by $Al_{0.3}Ga_{0.2}As$ barriers of thickness 15 nm. The procedure of the fabrication was described in detail in [9]. The middle regions of quantum wells (with thicknesses 5 nm) were doped as well as the middle region of the barriers with the same thicknesses. Consequently, the thickness of the undoped spacer layers from both sides of the barrier was 5 nm. We used Be as a p-type doping impurity introduced in concentration $1 \times 10^{17}$ cm$^{-3}$, the measured 2D concentration of holes at 300K was $1,5 \times 10^{11}$ cm$^{-2}$.

Figure 1 shows the temperature dependences of conductivity. It is seen that the latter corresponds to variable range hopping of the Mott type. For 2D it is described as

$$\sigma = \sigma_0 \exp(-T_0/T)^{1/3} \quad (1)$$

where

$$T_0 = C(N_f a^2)^{-1} \quad (2)$$

Here $N_f$ is the density of states at the Fermi level, $a$ is localization length; C = 13.8 being a numerical coefficient. For our samples $T_0$ was estimated as 1500 K.

Insert of fig.1 presents the results of magnetoresistance (MR) measurements obtained when the magnetic field normal to the plane of the structure was swept from -1 T up to +1 T with relatively fast sweep rate ($4 \times 10^{-3}$ T/s). It is seen that at weak fields MR is negative while at higher fields it tends to positive values. Such a behavior is consistent with theory of Shklovskii and Spivak [10] predicting for weak fields negative MR originating from interference of the different hopping trajectories including ones with intermediate underbarrier scattering. According to this theory

$$\ln\left(\frac{\sigma(H)}{\sigma(0)}\right) = \frac{1}{2\Phi_0} N\mu^2 a^2 \left(\frac{T_0}{T}\right)^{1/3} H \quad (3)$$

where $N$ is the concentration of scatterers, $\mu$ is a scattering amplitude and $\Phi_0$ is a magnetic flux quantum. On the other hand, at higher fields the contribution of wave function shrinkage positive magnetoresistance [10]

$$\ln\left(\frac{R(H)}{R(0)}\right) = K \frac{T_0}{T} \left(\frac{eHa^2}{ch}\right)^2 \quad (4)$$

(where K = 0.0028) starts do dominate which explains the observed behavior.

Note, however, that in our system there exists an additional mechanism of magnetoresistance first suggested in [11] and then studied in detail in [12]. This magnetoresistance is related to the fact that in our samples we deal with several types of localized states - the ones related to the upper Hubbard band originating due to acceptors within the well (A$^+$) and the states of the lower Hubbard band originated from the acceptors within the barrier (A$^-$) which still can attract the hole[12]. It is a partial suppression of hops between centers of different types due to spin correlations within the magnetic field which leads to the magnetoresistance mentioned above. In the weak fields the corresponding estimate gives [13]

$$\ln\left(\frac{R(H)}{R(0)}\right) = \frac{g_u + g_l}{(g_u + g_l)^2} \frac{2\mu g H}{T} \quad (5)$$

where $g_u$ is a density of states for the upper Hubbard band, $g_l$ - the corresponding density in the lower Hubbard band, $\mu$ is the Bohr magneton.

In this paper our main goal was studies of a response to slow variation of magnetic field. Again, the magnetic field was swept between -1 T and + 1 T, however at much slower different sweep rates. Fig. 2 presents the dependence of voltage on sigma contacts of sample (the current was constant, I=1 nA)on time for the sweep rate $1 \cdot 10^{-3}$ T/s (for different signs of the current).

It is clearly seen that for both signs of the current the resistance slowly relaxes to smaller values. This decrease is estimated 2 % for this sweep rate. Fig.3 presents the time dependence of the resistance corresponding to the moments H = 0. As it is seen, it can be fitted by a logarithmic law.

# 3. Discussion

Let us consider the hopping site with equilibrium energy $\varepsilon$ (referred to the Fermi level) coupled to some TLS with interlevel splitting in equilibrium being equal $E$. The coupling potential $U(R)$ where $R$ is a distance between the hopping sites is defined as difference between the coupling energy for the two TLS states corresponding to the upper and lower levels of the TLS. If one creates an excitation on this hopping site (the electron or a hole depending on the sign of $\varepsilon$) the TLS splitting is changed $E \to E + U$. If $U$ is negative and $|U| > E$, the TLS changes its state with respect to the equilibrium one. Correspondingly, the excitation energy is also changed: $\varepsilon + U$ and for electron excitation its energy is lowered. For the hole excitation the same effect will be if $U > 0$. As a result, the presence of TLS leads to a formation of the polaron gap around the Fermi level with a width $2U$ and the depth depending on the concentration of TLS. The form of the gap can be found as follows [6]. If the TLS density of states is $P(E, R, \tau) = \bar{P}/\tau$ where $\bar{P} = const$ then the distribution function of $U$ can be found from the following equation:

$$F(U)\mathrm{d}U = \int_{\tau_{min}}^{\tau_{max}} \mathrm{d}\tau \tau^{-1} \int_0^U \mathrm{d}E \bar{P} 2\pi R \mathrm{d}R \tag{6}$$

or

$$F(U)\mathrm{d}U = \ln \frac{\tau_{max}}{\tau_{min}} U \bar{P} 2\pi R \mathrm{d}R \tag{7}$$

The r.h.s. simply describes the probability to find TLS ensuring the value $U$ within the interval $\rm d U$ which can be switched between the smallest and largest time scales of the experiment. Thus the form of the polaron gap (which excludes all the states with energies $< U$) can be described as

$$\theta(U) = -2\pi \int_U^\infty \mathrm{d}U' \cdot U' \bar{P} R(U') \frac{\mathrm{d}R}{\mathrm{d}U'} \ln \frac{\tau_{max}}{\tau_{min}} \tag{8}$$

In our paper [6] we have considered electronic TLS formed from the pairs where one of the site is occupied while another is empty. The two states of the aggregate correspond to opposite direction of all of the "spins". The transition between the two states can be due to either multielectron hop, or due to motion of the "domain wall" separating parts of the aggregate with different phases of spin orientation. In this case the size of the aggregate is relatively large and coupling is most effective when the hopping site is close to one of the sites forming the aggregate. This coupling is naturally the Coulomb one. The coupling with the rest site of the aggregate is considered to be much weaker. Note that in this case the distribution function $\bar{P}$ should be multiplied by the number $N$ of the sites within the aggregate since the hopping site can be coupled to each site within the aggregate which increases the probability of the coupling. In what follows we will absorb this factor into $\bar{P}$. As it is seen, for the Coulomb potential $U = e^2/\kappa R$ the integrand in Eq.8 is $\propto (U')^{-2}$ and, correspondingly, the integral is controlled by the lower limit. The apparent divergency at $U \to 0$ has a cut-off related to the fact that the small values of interaction energy $U$ correspond to large distances between the hopping site and the aggregate. Thus the Coulomb coupling is replaced by much weaker dipolar coupling originating from (random) dipolar moment of the aggregate. So we assume that the cut-off is given by the energy $\varepsilon_h$ corresponding to the typical coupling energy between the sites separated by the typical hopping length $r_h$.

Correspondingly, we assume that the gaps with $U < \varepsilon_h$ (where $\varepsilon_h$ is the hopping energy band) can not effectively affect the hopping transport. As a result, the density of states near the Fermi level as a function of energy can be described by interpolation relation as

$$\frac{\delta g}{g} \sim 2\pi (e^2/\kappa)^2 \bar{P} \left( \frac{1}{(\varepsilon^2 + \varepsilon_h^2)^{1/2}} \right) \ln \frac{\tau_{max}}{\tau_{min}} \tag{9}$$

Here $\tau_{max}$ is the quenching time spent by the system before the application of the magnetic field while $\tau_{min}$ is the time of the following evolution.

However in our case the situation is some more complicated than the one considered in our paper [6]. Indeed, we deal not only with "occupied" and "empty" states which for the upper Hubbard band could be related to $A^+$ and $A^0$ centers. As it was mentioned above, some holes can be still captured by the acceptors in the barrier [12]. If $e^2/\kappa r + (\varepsilon_0 - U) < 2e^2/\kappa d$ where $\varepsilon_0$ is the acceptor Bohr energy, $U$ is the Hubbard energy, $r$ is a distance between the acceptor in the barrier and closest acceptor in the well then the hole stays to be attracted to the acceptor in the barrier. The complex can be destroyed by the reaction $\tilde{A}^0 + A^0 \to \tilde{A}^- + A^+$ As a result, at the Fermi level there coexist the states of the upper Hubbard band mentioned above and of the lower Hubbard band where the states below the Fermi level corresponds to $\tilde{A}^0$ states.

The application of the magnetic field $H$ does not affect energy of $A^+$ states with spin equal zero, but affects $A^0$ and $\tilde{A}^0$ states due to Zeeman energy $\mu_B g H$ where $\mu_B$ is the Bohr magneton. As a result, the holes redistribute between the sites and the chemical potential is shifted [12]:

$$\delta\mu = \mu_B g H \frac{g_u - g_l}{g_u + g_l} \qquad (10)$$

Thus an instant application of magnetic field leads to a shift of chemical potential. To estimate the corresponding change in conductance we have in mind that in our case $\mu_B g H \leq T$, and, correspondingly, $\delta\mu \leq T$ and in any case $\delta\mu \ll \varepsilon_h$. Thus the corresponding increase of the conductance is only quadratic in terms of $(\mu g H / \varepsilon_h)$ and directly follows the evolution of H which can not explain our results.

The more important effect of the magnetic field is related to the fact that a presence within the aggregate of the pairs of $A^+$ and $\tilde{A}^-$ states the magnetic field affects the interlevel spacing $E$ since in one of the states of the pair becomes neutral. If the charged pair corresponds to the upper state, magnetic field increases $E$ by an addition $N_1 \mu g H$, where $N_1$ is the number of the corresponding pairs in the aggregate. Thus the polaron gap created by this aggregate can be destroyed by the field if $E(H=0) + N_1 \mu g H > |U|$ since the second state becomes energetically preferable despite of a presence of the correlation energy $U$. One could assume that this effect is compensated by the TLS where the charged pair $A^+$, $\tilde{A}^-$ exists in the lower energy state since in this case magnetic field decreases $E$ and apparently allows new aggregates to join to the formation of the polaron gap. However, according to our consideration, it is the upper energy state which interacts with the corresponding hopping site creating the polaron gap. One sees that in this state the charged pair mentioned above becomes neutral and the sites entering the pair can not be participate in the polaron gap. Correspondingly, the symmetry between the aggregates with the charged pairs in the lower and in the upper states breaks and the contribution to the polaron effect of the TLS with charged pairs in the upper state is more strongly affected by the magnetic field. Since the number of the sites within the aggregate is not expected to be very large, for the simplicity we will assume that $N_1 = 1$. In this case the decrease of the number of TLS participating in the polaron effect can be estimated as

$$-\frac{g_l}{g_u + g_l} \frac{\mu g H}{\varepsilon_h} \ln\left(\frac{t_H}{t_{min}}\right) \qquad (11)$$

In Eq. 11 we have taken into account that the aggregates with $E(H=0) + N_1 \mu g H > |U|$ gradually decay to their "passive" state with no polaron effect according to the law $\ln(t_H / t_{min})$ where $t_H$ is the time when magnetic field is applied. The corresponding time behavior seems to hold until $t_H < t_{max}$. It is important that this effect is linear in $H$ and thus is expected to dominate over the quadratic effect mentioned above.

In our experiments we deal with AC magnetic field, however the effect in question does not depend on the sign of the magnetic field and we can put to our estimates some average value of $H$. One also notes that the shift of the chemical potential is relatively small (at least with respect to $\varepsilon_h$) and thus it does not significantly affects the existing polaron gaps. This fact explains experimentally observed gradual logarithmic increase of conductances at the moments of $t$ corresponding to both $H = 0$ and $H = H_{max}$.

As for the quantitative estimate, we shall know the TLS distribution function. For the case of the electronic TLS discussed above, one has a rough estimate (see [6])

$$\bar{P} = \frac{N exp - \lambda N}{(N\rho^2) N^{1/2} e^2 / \kappa \rho} \quad (12)$$

Here $N$ is the number of pairs of sites forming the aggregate, the exponent is a statistical factor describing the probability to construct the bistable aggregate, $\rho$ is the typical distance between the sites forming the aggregate. Here the first factor in the denominator describes the typical volume of the (2D) aggregate while the second - typical scatter of the TLS energy splitting. The factor $\lambda$ depends on the competition between the Coulomb interactions within the system and scatter of single-particle energies. Indeed, for weak Coulomb interactions the system is in its ground state and occupation of all single-particle states is given. It is the Coulomb correlations which allow to have metastable configurations with close total energies. Unfortunately $\lambda$ is not known and can depend on realization of the Coulomb glass. However one expects that large relaxation times are available at small $N$ and thus the exponential is not too small. If one assumes that it is around 0.1 while $\rho$ is about characteristic hopping length $\xi a \sim 100nm$, $N^{3/2} \sim 10$ one estimates $\bar{P} \sim 10^{23}$ cm$^{-2}$ erg$^{-1}$. Thus making use of the estimates of Eq.9, Eq.11, with an account that for the field direction normal to GaAs well $g \sim 1$ and assuming $g_l = g_u$ one obtains

$$\frac{\delta G}{G} \sim 0.01 \left( \ln \frac{t_H}{t_{min}} \right) \quad (13)$$

which by order of magnitude correlates with the experimental data.

Actually in our case the effect of magnetic field is to some extent similar to the effect of the gate voltage in the experiments [2-5]. However there are important differences. First, the direct effect of the variation of the gate voltage (related to its effect on DOS) is much stronger than any mechanism of magnetoresistance which apparently does not allow to visualize relatively small contribution of the TLS polarons in doped semiconductors. Second, relatively large sweep rate for the gate voltage implies significant nonequilibrium factor which also can lead to masking effects preventing observation of TLS contribution if the latter is small (which is the case of the Mott-type hopping, [6].

To conclude, we observed slow relaxation of magnetoresistance as a response to applied magnetic field in selectively doped p-GaAs-AlGaAs structures with partially filled upper Hubbard band. We explain this behavior as related to the properties of the Coulomb glass formed by the charged centers with an account of spin correlations which are sensitive to external magnetic field.

## 4. Acknowledgements

This work was supported financially by the Russian Foundation for Basic Research (project no. 06-02-17068).

**References**

[1] S. Kogan, Phys. Rev. B 57 (1998) 9736.
[2] M. Ben Chorin, Z. Ovadyahu and M. Pollak, Phys. Rev. B 48 (1993) 15025
[3] Z. Ovadyahu and M. Pollak, Phys. Rev. Lett., 79 (1997) 459. F. Gay, cond-mat/0701560 (unpublished).
[4] Z. Ovadyahu, Phys. Rev. B 73 (2006) 214208
[5] A. Vaknin, Z. Ovadyahu, and M. Pollak, Phys. Rev. B 65 (2002) 134208
[6] V.I.Kozub, Yu.M.Galperin, V.M.Vinokur, A.L.Burin arXiv:0805.3840
[7] Electron polarons in hopping systems were introduced in A. L. Efros, J. Phys. C9 (1976) 2021 and N. F. Mott, Phil. Mag. 34 (1976) 643; see also M. Pollak and Z. Ovadyahu, Phys. Stat. Sol. (c) 3 (2006) 283 .
[8] A. L. Burin, B. I. Shklovskii, V. I. Kozub, Y. M. Galperin,and V. Vinokur Phys. Rev. B 74 (2006) 075205
[9] N. V. Agrinskaya, V. I. Kozub, Yu. L. Ivanov, V. M. Ustinov, A. V. Chernyaev, and D. V. Shamshur JETP 93 (2001) 424
Translated from Zhurnal Eksperimentalino i Teoretichesko Fiziki, Vol. 120, 480-485.(2001)
[10] B. I. Shklovksii and B. Z. Spivak, in Hopping Transport in Solids, Ed. by M. Pollak and B. Shklovskii (Elsevier,Amsterdam, 1991), p. 271
[11] H.Kamimura, T.Takemori, A.Kurobe, in: Anderson localization, ed. Y.Nagaoka and H.Fukuyama, Springer Series in Solid state science,(Springer, Berlin, 1982) v.39 p.156 5).
[12] D.M.Larsen, Phys.Rev.B.47 (1993) 16333
[13] K.A. Matveev, L.I. Glazman, Penny Clarke, D. Ephron, M.R. Beasley, Phys. Rev. B 52 (1995) 5289

Fig.1. Temperature dependence of resistance. Insert: magnetoresistance trace at T = 1,4 K and sweeping rate  $4 \times 10^{-3}$ T/s.

Fig.2. Voltage on the sample as a function of time for the opposite directions of the current. The magnetic field sweeping rate is $1 \times 10^{-3}$ T/s. Arrows indicate the moments corresponding H = 0.

Fig.3. Time dependence of the sample resistance at the moments corresponding H = 0.

**Figure(s)**
**Click here to download high resolution image**

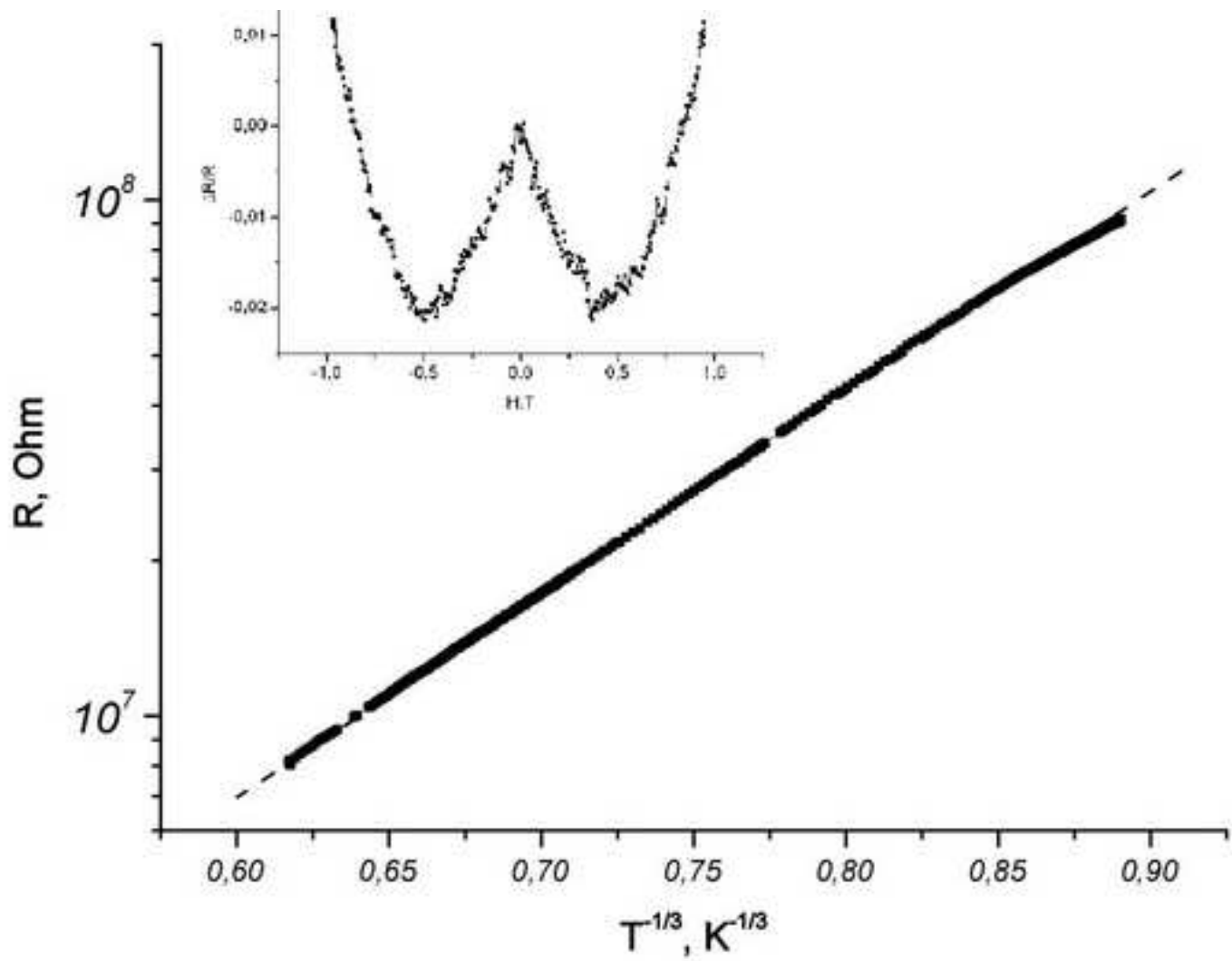

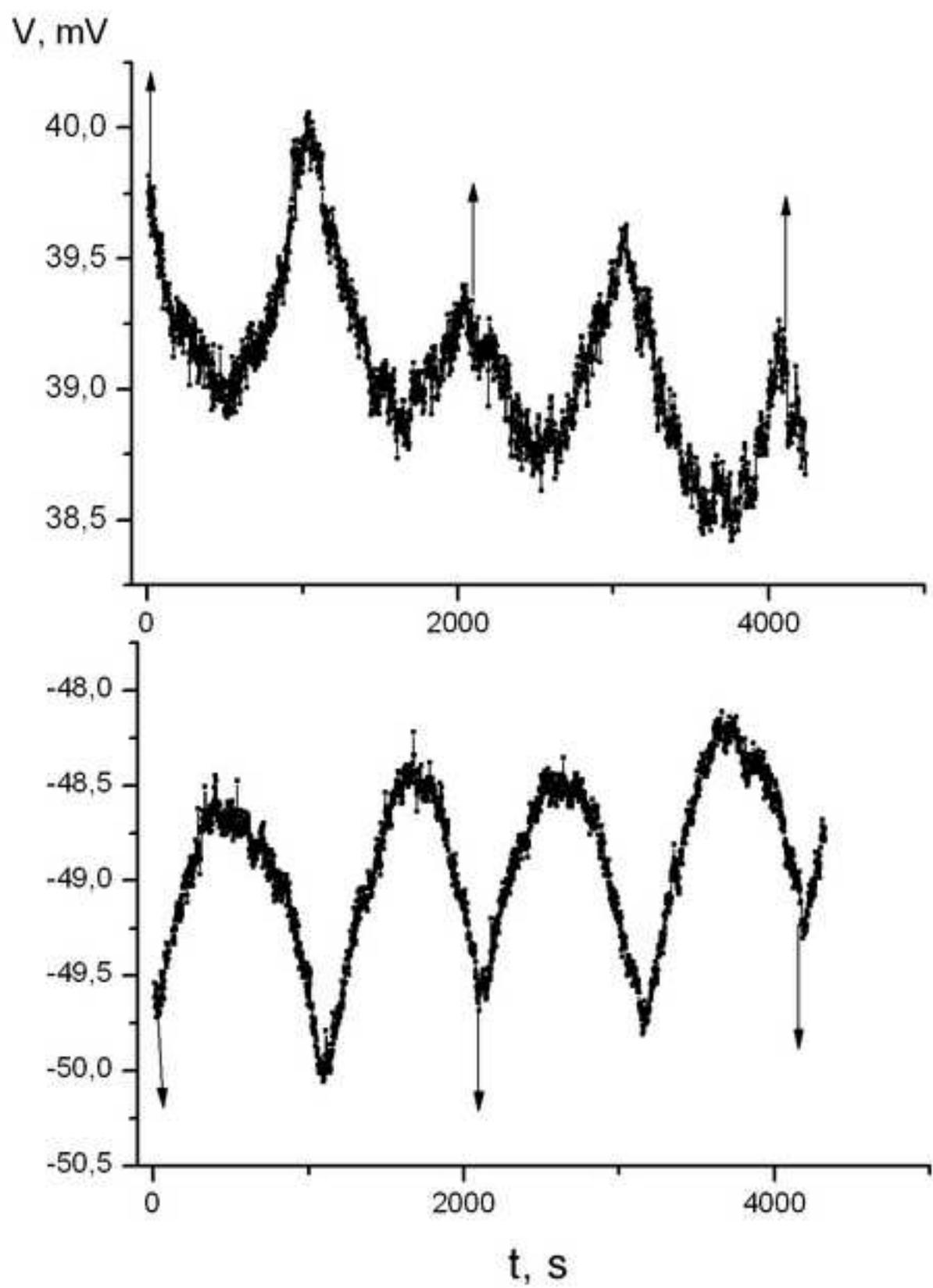

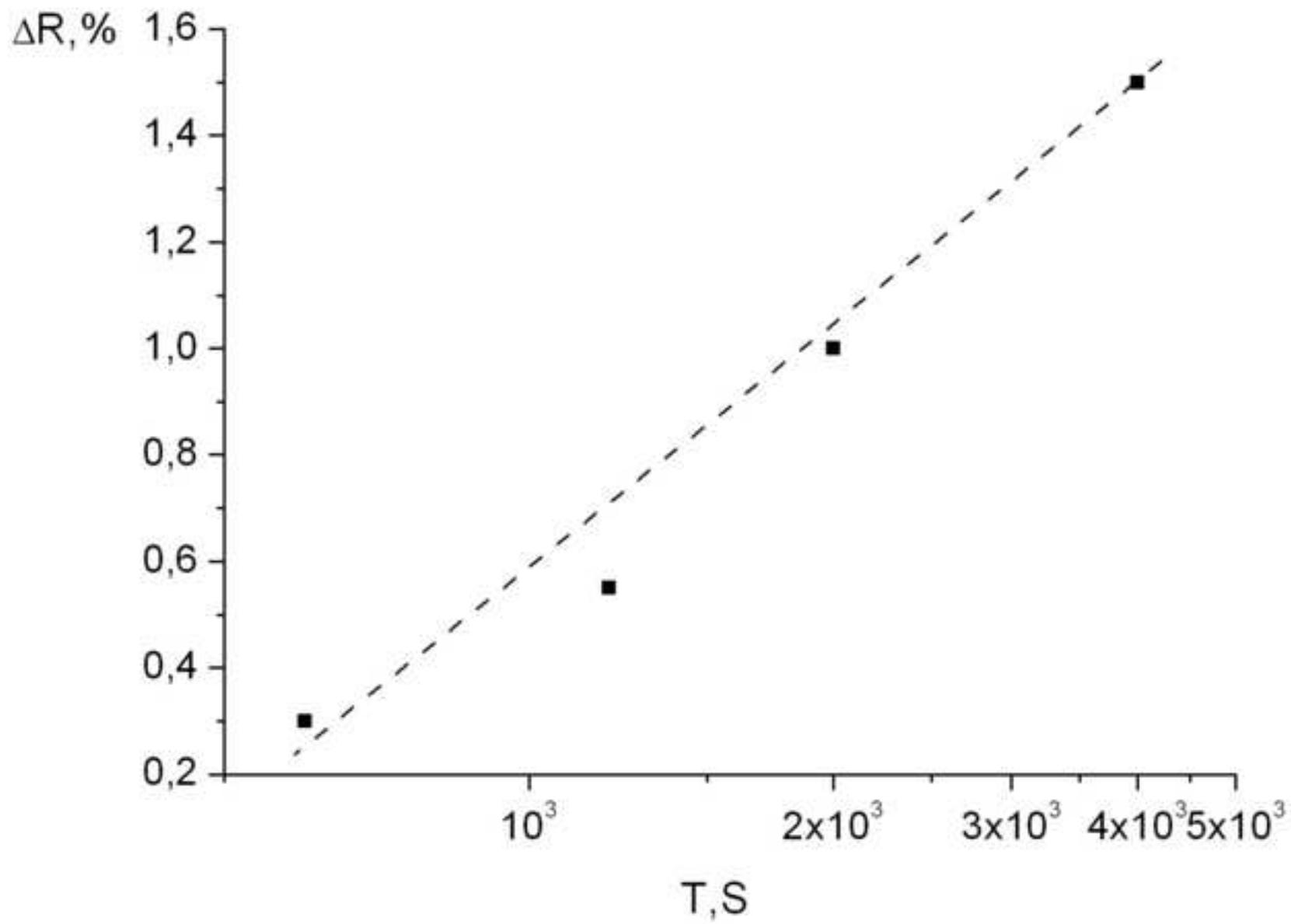

**\* Response to Reviewers**

Dear Reviewer,
We are grateful you for your remarks.
We have revised our manuscript on the base of these remarks.
1) We have included PACS numbers and keywords to the text.
2) We have rewritten the list of references according to your remarks.
3) We have made some corrections according to your remarks:
a) "occupation number due" - to "occupation numbers due"
 (abstract)
b) "but affect" - to "but affects" (p.4)
c) "being Bohr" - to "is the Bohr" (p.4)
Sincerely yours - N. Agrinskaya